# ArchGuard-APR: Architecture-Constraint-Aware Multi-Agent Program Repair Using Repository Graphs and Execution Feedback

**Majid Feiz Deilami**

M.Sc. Student in Software Engineering

Department of Computer Engineering,
Lahijan Branch, Islamic Azad University,
Lahijan, Iran

Corresponding author: majid.feiz@iau.ir

ORCID: 0009-0006-8439-1839

*Author manuscript — not peer reviewed*

## Abstract

Repository-level automated program repair must reason about more than a localized failing statement. A candidate change can satisfy a narrow test while violating dependency boundaries, modifying an unsafe path, or relying on context that omits relevant repository structure. ArchGuard-APR is an architecture-aware repair prototype that integrates deterministic repository analysis, immutable repository graphs, explicit architecture profiles, graph-guided context retrieval, role-based repair orchestration, patch-policy enforcement, isolated execution feedback, provider cost controls, and canonical source-free evidence. We evaluate the implemented pipeline through an exploratory paired feasibility study, not a confirmatory effectiveness experiment. Six executions were planned; four completed and two black-4 conditions ended as pre-provider infrastructure attrition. Across two valid paired tasks, baseline and ArchGuard conditions each achieved 0/2 successful repairs, and all four completed outputs were classified as invalid patches. The completed executions consumed 35,431 input tokens and 2,737 output tokens (38,168 total) at USD 0.1296325. No effectiveness advantage was observed, and the sample provides no useful inferential basis. The study instead establishes the engineering feasibility of composing architecture constraints with a reproducible repair pipeline and documents patch validity and infrastructure robustness as the principal targets for future work.

# 1. Introduction

Automated program repair (APR) seeks to generate modifications that correct faulty software. Modern repository-level tasks broaden the problem beyond a single function: tools must locate relevant code, respect project organization, generate an admissible patch, and validate behavior in a controlled environment. Language-model agents add flexible planning and tool use, but flexibility also increases the number of boundaries at which a run can become invalid. A plausible textual response is not necessarily a parseable patch; a patch that applies is not necessarily safe; a test-passing patch is not necessarily consistent with architectural constraints.

ArchGuard-APR treats these boundaries as explicit research objects. Its pipeline separates a vendor-neutral domain and application core from infrastructure adapters. Repository source is first converted into a validated snapshot. Graph construction consumes the snapshot rather than rereading the repository. Architecture evaluation consumes only a validated graph and an explicit, versioned profile. Context retrieval operates over bounded artifacts. Patch inspection and execution are gated by policy, isolation, budgets, and deterministic artifact persistence. This design favors auditability and fail-closed behavior over maximizing apparent repair yield.

The paper makes two different kinds of contribution. The engineering contribution is an implemented prototype integrating repository analysis, graphs, architecture constraints, retrieval, repair roles, patch validation, execution feedback, provider abstraction, and canonical artifacts. The scientific evidence is deliberately narrower: an exploratory evaluation demonstrates operational integration, records four invalid-patch outcomes, and exposes two infrastructure attritions. Engineering size, test count, and coverage are not treated as evidence of repair effectiveness.

The evaluation is negative and inconclusive with respect to effectiveness. It contains only two valid paired tasks, both from PySnooper. Neither condition produced a successful repair. Therefore, the manuscript neither claims an improvement nor performs meaningful significance testing. Its value lies in a traceable system account, a reproducible evidence package, and a candid analysis of where an architecture-constrained repair pipeline failed.

# 2. Background and Motivation

Generate-and-validate APR typically searches or synthesizes candidates and evaluates them against an available correctness signal. Repository-level settings complicate every stage. Fault-relevant context may cross module boundaries; build and test environments can fail independently of a candidate; and test suites provide incomplete specifications. Architecture introduces another correctness dimension: dependencies and ownership boundaries encode maintainability intent that behavioral tests may not observe.

Architecture conformance should not be confused with automatic architecture inference. ArchGuard-APR uses human-reviewed profiles containing explicit module assignments and rules. This choice limits automation but makes the evaluated constraints reviewable. A violation is meaningful only relative to that profile. The system consequently reports unassigned or ambiguous modules and fails closed rather than silently treating an incomplete mapping as valid.

Repository graphs provide a stable intermediate representation for structural queries and context selection. The graph is built from a validated snapshot and contains modules, definitions, imports, and containment relationships. This representation supports deterministic traversal and avoids an uncontrolled source reread during downstream evaluation. It is primarily static and cannot capture all runtime behavior, reflection, generated code, or environment-dependent coupling.

Execution feedback is also necessary but insufficient. Tests can reject candidates and expose regressions, yet sandbox startup, dependency acquisition, checkout, and evaluator configuration may fail before scientific execution. A rigorous experiment must separate such infrastructure attrition from repair outcomes. Phase 13 applies that distinction to black-4: its two planned conditions did not reach the provider and are not counted as model failures.

## 2.1 Repair pipelines and outcome semantics

Classical test-suite-based APR decomposes repair into fault localization, candidate construction, candidate ranking, and validation. GenProg searches transformations derived from existing program material and uses tests as its fitness signal [4]. SemFix instead derives repair constraints through semantic analysis and synthesis [5], while Angelix extends symbolic synthesis to scalable multiline changes [7]. Prophet learns a probabilistic model from successful human patches to rank generated candidates [8]. These systems differ substantially in representation and search, but all illustrate a central point: the mechanism that proposes a patch is distinct from the oracle that accepts it.

This distinction motivates precise outcome language. A *candidate* is any proposed edit. A *valid* candidate satisfies the transport, syntax, path, size, and policy contract required to enter evaluation. A *plausible* patch applies and satisfies the available test suite. A *correct* patch satisfies the intended behavior, including behavior not completely encoded by those tests. Large-scale Defects4J experiments found that test-adequate patches can remain incorrect [9]. ArchGuard-APR adds an architectural conformance dimension between admissibility and final success, but does not treat that dimension as a substitute for behavioral correctness.

These terms also separate scientific outcomes from infrastructure events. A provider response that cannot be parsed into an admissible patch is a completed scientific execution with an `invalid_patch` classification. A repository that cannot be acquired or validated before provider transmission has no provider outcome at all. Combining the two would incorrectly attribute acquisition reliability to the repair method. The Phase 13 accounting therefore keeps planned slots, completed calls, valid pairs, and attrited slots as separate quantities.

## 2.2 Repository-level repair and context

Method-level APR often assumes that a suspicious location and nearby code are already available. Repository-level issue resolution is broader. SWE-bench instances pair real issue descriptions with repository snapshots and tests; resolving them can require coordinated edits across files and interaction with an execution environment [20]. BugsInPy similarly emphasizes reproducible faults in real Python projects [14], whereas Defects4J provides controlled Java faults [6]. These benchmarks are not interchangeable. They embody different languages, dependency ecosystems, task construction processes, and environmental assumptions.

Repository context is therefore both a resource and a risk. Too little context can hide definitions, callers, configuration, or ownership boundaries required for a sound change. Too much context can exhaust budgets, distract generation, disclose irrelevant content, or make executions difficult to compare. Retrieval-augmented generation formalizes a general separation between parametric generation and retrieved evidence [15]. Program-specific work represents syntax and semantics as graphs [11] or incorporates data-flow edges into learned code representations [16]. ArchGuard-APR does not claim that its static repository graph is equivalent to either learned representation. It adopts the narrower engineering lesson that explicit relations can support bounded, auditable selection better than an unrecorded recursive source dump.

## 2.3 Architecture and conformance

Software architecture has long been described through elements, relationships, and rationale [1]. Reflexion models operationalize comparison between an engineer's high-level model and source-level relations [2]. ArchGuard-APR draws on this conformance tradition: a profile states module assignments and allowed dependencies, and an evaluator compares a validated repository graph against that stated model. The profile is an input authored and reviewed by people, not an automatically discovered truth.

Architectural correctness is also different from style preference. A prohibited dependency can encode an isolation, ownership, testability, or vendor-independence boundary. Conversely, an incomplete or incorrect profile can reject an appropriate implementation. For that reason, profile provenance, schema validation, deterministic compilation, and explicit unassigned modules matter as much as the individual rules. The study asks whether such constraints can be incorporated reliably into a repair pipeline; it does not test whether any particular profile is universally desirable.

# 3. Related Work

## 3.1 Generate-and-validate, semantic, and learning-based repair

The APR literature spans behavioral repair, state repair, multiple oracle types, and many repair operators [3]. Within test-suite-based behavioral repair, GenProg is a foundational search-based approach [4]. SemFix and Angelix represent semantics- and synthesis-oriented alternatives [5], [7]. Prophet learns to rank candidates using properties of correct code [8], and Elixir incorporates object-oriented context in candidate selection [10]. These approaches establish that candidate-space design, context, ranking, and validation all influence observed repairability.

Neural repair introduced learned translation as a candidate generator. CoCoNuT combines context-aware neural translation models in an ensemble [13]. Large pretrained models broaden the available patch distribution without eliminating the need for localization, validation, or tests [19]. ArchGuard-APR uses a provider abstraction rather than proposing a new learned model. Its research focus is the surrounding system: deterministic acquisition, architecture-aware retrieval, typed orchestration, policy gates, isolated evaluation, and evidence persistence.

Empirical studies caution against equating a passing test suite with correctness. The Astor/Defects4J study distinguishes test adequacy from manually assessed correctness [9]. RepairThemAll evaluates multiple Java tools and benchmarks at scale, documenting variation across tool/benchmark combinations and providing an experimentation framework [12]. These observations inform ArchGuard-APR's explicit terminal classes and its refusal to reinterpret malformed output as a scientific success.

## 3.2 Benchmarks and repository-level issue resolution

Defects4J provides reproducible Java faults for controlled testing research [6]. BugsInPy provides a corresponding curated resource for Python projects [14]. SWE-bench changes the unit of work from a localized fault to a real repository issue and its execution environment [20]. Benchmark scale alone does not make evaluations comparable: qualification rules, versions, dependencies, tests, leakage controls, and success criteria remain consequential.

AutoCodeRover applies structured program search to repository issue resolution [22]. SpecRover adds intent extraction and a reviewer role [23]. RepairAgent uses an autonomous tool-using repair loop [21]. These systems motivate repository search, tool use, and feedback, but their reported studies use different task populations and protocols. ArchGuard-APR therefore does not compare its four completed executions numerically with their results. Its evidence supports only an internal baseline-versus- architecture-context description under one frozen protocol.

## 3.3 Agents, retrieval, and program graphs

ReAct interleaves language-model reasoning with environment actions [17], and Reflexion retains verbal feedback across trials [18]. Those general agent frameworks help explain why a repair workflow may benefit from explicit tools and feedback. They do not establish that adding more roles improves repair, nor do they prescribe the trust boundaries needed for untrusted repositories. ArchGuard-APR uses role-based services and typed state transitions; the term “multi-agent” denotes that orchestration structure rather than a claim about emergent collaboration.

Retrieval-augmented generation separates retrieved evidence from a generator's parameters [15]. Learning-to-represent-programs work shows how syntax and semantic relations can form graph inputs [11], while GraphCodeBERT uses data-flow information during representation learning [16]. ArchGuard-APR's repository graph is neither a neural embedding nor a complete program-dependence graph. It is a canonical structural artifact used for deterministic queries, architecture evaluation, and context provenance.

### 3.4 Architecture-aware positioning

Perry and Wolf provide a foundational account of software architecture [1], and software reflexion models compare high-level expectations with source structure [2]. ArchGuard-APR connects that conformance concern to repair orchestration. The verified sources do not justify a priority claim, and this manuscript makes none. The defensible positioning is an integrated prototype and experience report: explicit architecture profiles affect context and candidate validation, while source-free artifacts expose how the workflow reached each terminal state.

Table 10 compares only features established in the inspected primary sources. `NR` means that the feature was not reported or was not established during verification; it must not be read as evidence that a system lacks the feature.

## 4. Research Questions

**RQ1.** Can architecture constraints, repository graphs, context retrieval, and execution feedback be integrated into a deterministic and reproducible APR pipeline?

**RQ2.** What operational and infrastructure limitations emerge when applying the prototype to qualified real defects?

**RQ3.** What repair outcomes, token consumption, and provider costs are observed in the exploratory paired evaluation?

The questions separate integration feasibility from repair effectiveness. None asserts superiority. Their operational definitions and evidence sources are summarized in Table 2.

# 5. ArchGuard-APR Design

Figure 1 summarizes the implemented architecture, and Table 1 maps the major components to their roles and evidence.

Entrypoints → Application services → Vendor-neutral domain → Validated snapshots → Immutable repository graph → Architecture profile evaluation → Secure patch validation → Sandboxed execution feedback → Provider abstraction and cost guards → Canonical source-free artifacts

*Figure 1. ArchGuard-APR system architecture; components are implementation boundaries, not effectiveness claims.*

Entrypoints translate CLI or API requests into application services. Application services coordinate domain objects and ports while remaining independent of provider SDKs. Infrastructure adapters handle filesystem artifacts, Git worktrees, container execution, and model providers. This dependency direction is continuously checked by an architecture profile.

The repair workflow uses explicit roles and state transitions rather than an unrestricted conversation among agents. Deterministic fake transports test orchestration without making model calls. The live provider remains behind a port, and scientific execution requires explicit acknowledgements, frozen configuration, and cost gates. Thus “multi-agent” describes implemented role separation and coordination; it does not imply that the exploratory study validated autonomous collaboration.

## 5.1 Component responsibilities, trust, and data flow

The entrypoint layer translates operator intent into typed application requests. It parses options, reports results, and selects adapters, but does not decide whether a patch is scientifically successful. Application services sequence repository analysis, graph construction, context compilation, repair orchestration, and evidence finalization. Domain objects define identifiers, validation rules, state transitions, and terminal classifications. Infrastructure implements Git, filesystem persistence, isolated execution, and provider transport. This direction permits domain classification to be tested with fake providers and temporary repositories while live credentials remain confined to an adapter.

Frozen protocol, population, pricing, and fingerprints become trusted control data only after schema and identity checks. Target source, issue text, generated patches, tool output, and archives are untrusted payloads crossing explicit validation boundaries. Source-free diagnostics retain categories and stable fingerprints while omitting raw source, prompts, responses, environment values, and secret-bearing paths.

The principal data flow is monotonic. Acquisition produces an immutable checkout identity; analysis produces a canonical snapshot; graph construction consumes the snapshot rather than the checkout; constraint evaluation consumes the graph and compiled profile; and context compilation records a bounded selection manifest. The repair loop then produces a typed candidate result. Inspection either rejects that result or yields an admissible patch for isolated evaluation. Each transition persists evidence only after validation.

Canonicalization does not make an external model deterministic. It makes inputs, classifications, aggregates, and source-free outputs comparable. Maps use stable key ordering, unordered collections become stable sequences, and hashes cover exact serialized bytes. Artifact identities exclude temporary checkout roots, allowing repeated offline builds in different directories to be compared.

# 6. Repository Analysis and Graph Construction

Repository analysis parses Python files without importing or executing target code. It records stable identities, module and definition summaries, imports, diagnostics, and a repository fingerprint. Analysis output is persisted canonically. Failed or skipped files are explicit rather than silently discarded.

Graph construction accepts only a validated snapshot. Nodes describe repositories, files, modules, classes, and functions; edges capture containment, definitions, and import relationships. Resolution distinguishes internal, external, ambiguous, and unresolved imports. Graph construction never rereads target source, which makes the graph a reproducible derivative of the snapshot rather than a moving view of a worktree.

The representation is intentionally bounded. It is not a complete semantic program graph and does not model arbitrary dynamic behavior. Its purpose is stable structural querying, architecture evaluation, and retrieval support.

### 6.1 Snapshot and graph invariants

Analysis walks only paths admitted by repository policy. It rejects traversal outside the protected root, symbolic-link escapes, unsafe file types, and outputs located inside the analyzed tree. Python files are parsed as text into an abstract syntax representation; they are not imported and module initializers are not executed. Parse failures become explicit diagnostics. Snapshot summaries record discovered, parsed, failed, and skipped files so downstream users can assess completeness.

Stable identifiers derive from normalized repository-relative locations and qualified definitions, not temporary checkout roots. Graph nodes represent repositories, files, modules, classes, and functions. Containment and definition edges preserve hierarchy; import edges record dependencies. Resolution labels imports internal, external, unresolved, or ambiguous instead of forcing a guessed target. Treating unresolved imports as absent would understate coupling, while inventing an internal edge could create a false violation.

Graph builders validate the snapshot fingerprint and operate solely on snapshot data. Query services provide bounded traversals, dependency neighborhoods, cycles, and structural summaries. The design trades some whole-program precision for reproducibility and a narrower trust surface. Generated code, reflection, monkey patching, native extensions, and runtime dependency injection may remain outside the representation and are documented as validity limits.

## 7. Architecture Constraint Model

Architecture profiles are explicit YAML documents with versioned schemas. Selectors assign modules to components; rules constrain permitted dependencies, cycles, and selected external frameworks. Compilation separates YAML loading from domain validation. Evaluation operates on validated graphs and emits a source-free report with profile and graph fingerprints.

Unassigned and ambiguous modules can invalidate a report. This fail-closed behavior matters because evaluating only a subset of modules could create false confidence. Profiles require human judgment: component boundaries and allowed dependencies represent intended design, not facts inferred from the repository. Consequently, the system can test conformance to a reviewed model but cannot establish that the model is uniquely correct.

## 7.1 Profile lifecycle and rule interpretation

A profile begins as YAML authored against a versioned schema. Loading performs data conversion; compilation performs semantic checks for unique components, valid selectors, known rule endpoints, and deterministic priority. The compiled profile receives a fingerprint and is the only form consumed by evaluation. Separating these stages prevents parser behavior from becoming hidden constraint semantics.

Profiles are reviewed and frozen before execution. Changing a profile after observing outcomes could alter the treatment opportunistically, so Phase 13 couples profile and protocol fingerprints to its checkpoint and final manifest. The same fail-closed identity rule applies to population order, provider configuration, and model. A mismatch stops rather than silently updating stored provenance.

Selectors map modules to components. Evaluation reports ambiguity when incompatible selectors claim a module and incompleteness when required modules remain unassigned. Dependency rules compare resolved internal edges against allowed component relations; cycle rules operate on the validated graph; external-dependency rules constrain selected frameworks without pretending every third-party import is internal. Reports preserve severity, evaluated and skipped rule counts, graph/profile fingerprints, and stable violation identities.

# 8. Multi-Agent Repair Workflow

The workflow coordinates analysis, retrieval, generation, validation, and evaluation roles through typed requests and results. Figure 2 shows the main sequence.

Validated repository snapshot → Graph and context retrieval → Architecture constraints → Provider request under budget gate → Patch parsing and security policy → Sandboxed evaluator → Terminal invalid-patch record → Source-free execution result

*Figure 2. End-to-end repair workflow with fail-closed validation gates.*

Provider requests occur only after configuration and budget checks. Responses are converted into patch candidates, then subjected to parsing, path, size, and policy checks. Invalid candidates become terminal records rather than being coerced into plausible patches. Valid candidates may proceed to an isolated evaluator. Retries are bounded and accounted for.

This design makes failure provenance visible. Provider failures, invalid patches, evaluator outcomes, budget stops, and infrastructure failures are distinct states. That separation is essential for the Phase 13 result: the four completed calls yielded invalid patches, while black-4 stopped during checkout before a call.

## 8.1 Candidate lifecycle and accounting

A generation request binds task identity, condition, context-manifest fingerprint, provider configuration fingerprint, and budget reservation. The adapter returns transport metadata, usage, and content to a parser boundary. Accounting marks calls started, completed, and carrying usage independently so interruption cannot make a charged call disappear or be counted twice on resume.

Candidate parsing recognizes only the frozen patch envelope. Inspection does not normalize unsafe paths or malformed syntax into a favorable candidate. It checks size ceilings, repository-relative targets, prohibited files, link behavior, and modification policy before applying anything. An output that a human could repair manually remains invalid in the experiment. This avoids an unrecorded human postprocessor and keeps handling identical across conditions.

An admissible patch is applied to a disposable worktree at the frozen revision. The evaluator runs a frozen command under time, resource, network, mount, and output constraints. Test success is necessary but may be followed by architecture and policy checks. Cleanup is a terminal obligation: incomplete cleanup is fatal rather than discarded when a candidate otherwise appears successful.

### 8.2 Roles and provider abstraction

Analysis, retrieval, generation, validation, and evaluation have different inputs and trust levels. They communicate through typed records rather than a shared mutable conversation. Offline tests use deterministic transports that record invocations and return fixtures. Only the live adapter interprets credentials and provider-specific wire formats. This limits the meaning of “agentic” here. Prior work demonstrates action/feedback loops [17], [18] and repair-specific tool use [21]–[23]; ArchGuard-APR contributes auditable orchestration, not evidence that role decomposition improves patches.

## 9. Context Retrieval and Execution Feedback

Context retrieval combines explicit problem information with bounded graph queries. It selects stable module and symbol neighborhoods and emits a manifest fingerprint. Budgets prevent repository content from expanding without bound. Architecture treatment adds reviewed constraint information without changing the underlying task order or evaluator.

Execution feedback is produced in a restricted container configuration with resource, path, network, and output controls. The evaluator distinguishes candidate behavior from infrastructure preparation. Isolation reduces risk but does not eliminate it: container runtimes, dependencies, kernels, and host configuration remain part of the trusted computing base.

### 9.1 Retrieval budgets and provenance

A context request begins with explicit problem information, traceback data when available, and stable requested symbols or modules. Graph queries expand only within configured depth and item budgets. The compiler orders selected items deterministically and records why each was included. If a limit is reached, truncation appears in the manifest rather than being hidden in prompt assembly.

The architecture condition adds reviewed component assignments, applicable dependency rules, and violations associated with the selected neighborhood. Baseline omits that treatment while retaining the same task identity and orchestration. This contrast is narrower than comparing two different repair systems. It isolates configured architecture context in principle, but the realized sample is too small and unsuccessful to estimate its effect.

### 9.2 Sandboxed feedback

Execution plans identify an immutable image, command, worktree mount, timeout, memory and CPU limits, network policy, and output ceilings. The evaluator captures normalized exit status and bounded output metadata. Candidate worktrees are temporary and never become a new analysis source; structural checks continue from validated graph artifacts.

Sandboxing is defense in depth, not proof of containment. Image provenance, runtime implementation, kernel vulnerabilities, dependency installers, and host configuration remain threats. The workflow therefore distinguishes setup or checkout failures from test failures and retains cleanup state. A future deployment should combine minimal images, locked dependencies, least-privilege mounts, network denial, resource quotas, and independent host monitoring.

## 10. Security, Isolation, and Reproducibility

The implementation rejects unsafe paths, links, oversized artifacts, unexpected archive members, identity drift, dirty worktrees, and source/gold leakage. Provider credentials remain adapter-scoped. Cost state tracks calls started, completed, and carrying usage so that partial runs cannot silently double-count or lose expenditure.

Canonical JSON, stable ordering, content fingerprints, and source-free diagnostics support artifact comparison. The final source artifact SHA-256 is `48080661ddbff3e486d47cebc63754e29d4dd4664febbb7f8143484e4c03e99d`; the final result fingerprint is `761d9853f6ecfb4dafdaad81de781c4d49108b8b82912aeec509e173c7d56c70`. These identities support integrity checks, not independent replication of stochastic provider output.

# 11. Experimental Methodology

Phase 13 is an exploratory feasibility evaluation with a frozen population, ordered schedule, provider configuration, model snapshot, prices, and budget. Three qualified tasks each had baseline and ArchGuard conditions, producing six planned execution slots. The treatment changed the supplied architecture context while retaining the fixed orchestration and validation policy. Table 3 records the frozen population, while Table 7 makes the analysis denominator explicit.

The primary repair outcome was a valid evaluated repair, but terminal classifications preserve earlier failure points. Four PySnooper executions reached the provider. The two black-4 executions did not: ordinal 5 encountered checkout `validation_failure` with diagnostic fingerprint `71f60e3969313f139c493cad31ee20640fd6c2870d70411debdd5ab1956994e1`, and the paired ordinal was closed as the same task-level infrastructure attrition. Figure 3 presents the accounting.

6 planned executions → 4 completed provider outcomes → 2 black-4 pre-provider attritions → 2 valid paired PySnooper tasks → Baseline 0/2; ArchGuard 0/2 → Descriptive exploratory conclusion only

*Figure 3. Experimental protocol and completed/attrited accounting.*

The analysis is descriptive. Two valid pairs and zero discordant successful pairs cannot support a useful inferential effectiveness conclusion. No significance test is reported.

## 11.1 Outcome-blind task qualification

Qualification preceded provider execution and did not inspect model-generated outcomes. Candidate tasks had to provide an immutable buggy revision, stable upstream identity, reproducible problem description, applicable language/tool support, and leakage-safe acquisition. Qualification evidence records why each task entered the frozen population. This prevents later selection based on whether a condition appeared promising, although it cannot make three selected tasks representative of APR in general.

The population order is part of the protocol. Each task contributes a baseline slot followed by an ArchGuard slot, yielding six planned executions. Ordinals are stable across checkpoints so a resumed run can preserve completed evidence and continue at the next slot. A task-level pre-provider failure can close both paired slots as attrition when neither condition can be executed fairly.

## 11.2 Treatment construction and freeze

Before execution, repository snapshots and graphs were produced through the validated pipeline. Architecture profiles were human reviewed, compiled, and fingerprinted. Baseline and ArchGuard shared the task, provider configuration, model, repair orchestration, patch policy, evaluator, and budget logic. The intended treatment difference was the architecture-aware context supplied to the repair workflow. This design does not isolate every possible mediation effect, because architecture context can change input length and generated content, but it avoids confounding the treatment with a separate repair engine.

Population, protocol, and provider-configuration fingerprints were frozen and checked at start and resume. The model was `gpt-5.4-2026-03-05`. Provider prices and a USD 4.50 ceiling were fixed before execution. A call required explicit acknowledgement of provider cost and source processing. The runner reserved worst-case cost before transmission and reconciled reservation against reported usage after completion.

## 11.3 Classification and attrition policy

The result model separates preparation, provider transport, patch validity, application, evaluation, architecture conformance, and cleanup. A terminal scientific outcome requires that the provider call occurred and that its result reached a defined terminal class. Infrastructure exceptions before the provider boundary are not repair failures. Fatal authentication, billing, quota, fingerprint, security, corruption, budget, and cleanup conditions remain fail-closed.

Ordinal 5 (`black-4`, baseline) failed during checkout validation before provider use. The source-free diagnostic records stage `checkout`, category `validation_failure`, and fingerprint `71f60e3969313f139c493cad31ee20640fd6c2870d70411debdd5ab1956994e1`. Because the repository could not be prepared under the frozen acquisition policy, ordinal 6 was closed as paired task-level infrastructure attrition. Neither row receives a patch classification, token count, cost, or provider outcome.

## 11.4 Reproducibility controls

Checkpoint validation binds carried executions, call identifiers, usage, cost, fingerprints, and cleanup state. Resume policy distinguishes a historical stop reason from the active state of a new continuation. Completed ordinals cannot be scheduled again. Source-free checkpoints expose carried ordinals, next ordinal, accounting totals, historical diagnostics, and whether continuation is allowed.

The final import validated an artifact SHA-256 of `48080661ddbff3e486d47cebc63754e29d4dd4664febbb7f8143484e4c03e99d`. Canonical result builders then produced per-execution, paired, cost, token, attrition, and threats records. Tests recompute aggregates from rows rather than trusting prose. The final scientific result fingerprint is repeated in the manuscript and checked against the frozen Phase 13 package.

# 12. Exploratory Results

Four of six planned executions completed. PySnooper-2 and PySnooper-3 formed two valid pairs from one project. Baseline produced 0/2 successful repairs; ArchGuard produced 0/2. All four completed outputs were classified `invalid_patch`. Therefore, the observed success difference is zero, but this should not be interpreted as equivalence.

Table 4 contains the per-execution records and Table 5 the paired comparison. The two black-4 rows are infrastructure attrition and have no provider outcome, token usage, or repair classification.

Completed invalid patches: 4 | Infrastructure attrition: 2

*Figure 4. Exploratory execution accounting: four invalid patches and two infrastructure attritions.*

Baseline calls consumed 7,821 input and 1,037 output tokens at USD 0.0351075. ArchGuard calls consumed 27,610 input and 1,700 output tokens at USD 0.094525. Combined usage was 35,431 input and 2,737 output tokens (38,168 total), costing USD 0.1296325. Four provider calls completed, no retry occurred, reserved cost was zero, and USD 4.3703675 remained. In this tiny sample, ArchGuard used more input tokens and cost; no broader cost claim is justified. Table 6 provides the condition-level accounting.

Baseline: USD 0.0351075 | ArchGuard: USD 0.094525

*Figure 5. Observed provider cost by condition in the exploratory sample.*

Baseline: 7,821 input tokens | ArchGuard: 27,610 input tokens

*Figure 6. Observed input-token usage by condition in the exploratory sample.*

## 12.1 Descriptive interpretation

The correct denominator depends on the question. Operational completion is 4/6 planned slots, with 2/6 (33.3%) infrastructure attrition. Repair effectiveness is described over two valid paired tasks: baseline 0/2 and ArchGuard 0/2. Patch-format/policy yield is 0/4 admissible completed provider outputs, because all four were `invalid_patch`. Provider-call accounting is four completed calls with usage and zero retries. None of these denominators should be substituted for another.

The paired success difference is arithmetically zero, but the data do not establish equivalence. There are no discordant successful pairs and only one project contributes valid results. A hypothesis test would add ceremony without useful resolution. Confidence intervals would likewise be dominated by sample size and do not remedy missing project diversity. We therefore report exact counts and condition-level resource use only.

Observed input usage was substantially larger in the ArchGuard condition (27,610 versus 7,821 tokens), while output usage differed less (1,700 versus 1,037). This pattern is compatible with the architecture treatment adding context, but four calls cannot isolate treatment content from task-specific generation behavior. The cost difference is a property of these calls under frozen prices, not a forecast for future tasks or models.

## 13. Failure Analysis

The stored evidence establishes that all four completed responses failed patch validity. It does not establish one unique cause. Possible mechanisms include response formatting, incompatibility with the accepted patch grammar, path or policy rejection, insufficient task context, model behavior, or an integration mismatch. Because raw prompts and responses are intentionally absent from the source-free artifact, attributing individual failures among these possibilities would be speculation.

Strict validation has a dual effect. It protects repositories from malformed or unsafe modifications and preserves a clear experimental denominator. It can also lower apparent yield relative to systems that repair or reinterpret malformed outputs. The appropriate conclusion is not that validation caused the model failures, but that patch admissibility is a prerequisite the observed outputs did not meet.

Black-4 is categorically different. Its checkout validation failure occurred before provider transmission. It reveals an infrastructure robustness problem and reduces the evaluable population, but it supplies no evidence about baseline or ArchGuard repair ability on that task.

Planned: 6 → Completed: 4 → Infrastructure attrition: 2 → Invalid patch: 4 → checkout validation_failure → No provider outcome; no repair failure

*Figure 7. Attrition accounting preserves the planned denominator while excluding pre-provider failures from repair outcomes.*

## 13.1 Invalid-patch boundary

`invalid_patch` is intentionally a boundary classification, not a causal diagnosis. It says that a completed provider response did not satisfy the frozen candidate contract. The source-free evidence does not retain raw responses, so it cannot distinguish malformed diff syntax, an unsupported wrapper, an unsafe path, an excessive edit, or another policy violation for each ordinal. It would be inaccurate to attribute all four cases to the model, to prompt construction, or to the parser alone.

This uncertainty suggests a safe diagnostic refinement for future work. A validator can record a coarse source-free category such as envelope, syntax, target path, size, prohibited file, or application failure, plus a salted or canonical fingerprint, without retaining generated source. Such categories must be defined before the next study and tested for stability. Retrospectively relabeling Phase 13 responses would require unavailable payloads and would compromise the frozen result set.

Strict policy has a real tradeoff. Permissive normalization might convert some malformed text into an applicable patch, increasing apparent yield, but it introduces another repair operator whose behavior must be specified and evaluated. Manual cleanup would be even less reproducible. ArchGuard-APR chooses the conservative boundary: the provider must produce the accepted representation directly. A later protocol could prospectively compare strict parsing with a deterministic normalizer, but Phase 13 does not provide that evidence.

## 13.2 Checkout attrition

The black-4 diagnostic establishes the stage and class of failure while deliberately omitting the raw repository path and source. Checkout validation includes revision identity, upstream identity, tree shape, links, gitlinks, dirtiness, and leakage controls. The final study did not weaken those checks to obtain a larger denominator. That decision protects provenance but leaves black-4 scientifically unobserved.

This attrition demonstrates why infrastructure readiness should be exercised against every frozen task before paid execution. A source-free preflight can verify acquisition and cleanup without transmitting prompts. Such a preflight must itself avoid reading hidden answers into generated context and must not mutate the frozen checkout. Future protocols should make successful per-task acquisition a freeze prerequisite while preserving a separate audit trail for qualification failures.

# 14. Discussion

RQ1 receives engineering support: the repository contains an integrated, deterministic pipeline and canonical artifacts exercising its boundaries. This is integration feasibility, not demonstrated repair effectiveness. RQ2 identifies patch admissibility and checkout robustness as immediate limits. RQ3 records zero successes under both conditions and higher observed ArchGuard token use and cost in the two valid pairs.

The negative outcome redirects future work. Before expanding evaluation scale, the project should add source-free diagnostics capable of distinguishing patch syntax from policy categories, strengthen offline transport tests for admissible patches, and eliminate checkout attrition without weakening identity or leakage controls. A later study would require more tasks and projects, a prospectively defined analysis, and sufficient successful/discordant outcomes.

## 14.1 What the implementation establishes

The repository and its deterministic tests establish that architecture profiles, validated repository graphs, bounded retrieval, provider-neutral orchestration, patch inspection, isolated evaluation, cost accounting, resume checkpoints, and canonical evidence can be connected behind explicit ports. They also establish that offline fake transports can exercise provider boundaries without network access and that scientific facts can be rebuilt into publication artifacts.

They do not establish that the architecture context is useful to a model, that the selected graph is the best representation, that multi-role orchestration is preferable to a single loop, or that the security boundary resists every adversary. Most importantly, four invalid outputs do not estimate repair effectiveness. Engineering completeness and test coverage are evidence about implementation discipline, not about patch quality.

## 14.2 Architecture context and cost

Architecture information consumes context budget. It may guide a generator away from dependency violations, but it may also displace task-local evidence or add irrelevant rules. A useful retrieval policy should therefore expose the marginal items selected, their provenance, and the budget consumed. Future evaluation should record context composition by category and test whether architecture content changes admissibility, behavioral success, or conformance among otherwise valid patches.

The present cost observation illustrates rather than resolves this tradeoff. ArchGuard used more input tokens and cost in two paired tasks, but neither condition produced a valid patch. It is impossible to separate fixed treatment overhead from task-specific response behavior at this scale. Future studies should predefine resource outcomes alongside repair outcomes and report cost per attempted, valid, plausible, and correct patch.

## 14.3 Lessons for reproducible APR studies

First, a protocol needs outcome classes spanning the entire pipeline, not only pass/fail tests. Second, resumable paid studies require exactly-once call and cost accounting. Third, benchmark qualification must include acquisition, cleanup, and environmental readiness without examining repair outcomes. Fourth, negative outputs and attrition must remain visible; silently dropping either biases the study. Finally, claims should follow the realized denominator. A frozen plan of six does not become six completed executions when two slots never cross the provider boundary.

# 15. Threats to Validity

**Construct validity.** Success requires a patch to pass parsing, policy, application, architecture, and evaluator gates. This is intentionally strict but may differ from other APR definitions.

**Internal validity.** Frozen ordering and canonical artifacts reduce hidden drift, yet model behavior, prompt construction, integration code, and environmental state may influence outcomes. The evidence does not isolate causes of invalid patches.

**External validity.** Only two valid pairs and one represented project contribute outcomes. No generalization to other repositories, languages, defects, models, or providers is supported.

**Conclusion validity.** There were no successes and zero discordant successful pairs. The study is underpowered and supports no inferential or superiority conclusion.

**Infrastructure validity.** Two of six slots (33.3%) attrited at checkout. Separating those slots prevents outcome contamination but leaves the intended black-4 comparison unobserved.

**Model/provider dependence.** Results depend on `gpt-5.4-2026-03-05`, the frozen configuration, prices, and execution date. Future configurations may behave differently.

**Task selection.** Qualification was outcome-blind, but the small population and valid-result loss limit representativeness. **Patch policy** may reject outputs that another pipeline normalizes.

**Repository representation.** Analysis is primarily static and may miss runtime or reflective relationships. **Architecture profiles** encode human judgment and may be incomplete or contestable.

**Benchmark selection and maturation.** Defects4J [6], BugsInPy [14], and SWE-bench [20] demonstrate different controlled and repository-level benchmark designs. Phase 13 used a separately qualified three-task population. Results cannot be transferred to those benchmarks or to the wider bug population. Dependency aging and immutable-revision availability can affect whether historical tasks remain executable.

**Measurement and classification.** The `invalid_patch` class is reproducible but coarse. It protects against unsupported causal claims while limiting diagnostic resolution. Success requires multiple gates, so comparisons with work using test adequacy alone would need harmonized definitions.

**Human-authored profiles.** Review reduces accidental errors but does not eliminate subjectivity. Different maintainers may choose different components, selectors, severities, or allowed dependencies. A future study should measure inter-reviewer agreement and publish source-free profile rationale where repository confidentiality permits.

**Security.** Sandboxing and fail-closed boundaries reduce exposure but cannot eliminate risks from the runtime, dependencies, host, or malicious inputs. **Reproducibility** preserves inputs, aggregates, and source-free evidence, but cannot guarantee byte-identical stochastic provider responses. The consolidated assessment appears in Table 8.

# 16. Ethical and Reproducibility Considerations

No benchmark source, gold patch, credential, prompt, or provider response is included in the paper package. Negative results and attrition are retained. Costs and calls are reported rather than hidden. The manuscript uses only verified citations and marks unresolved human decisions. Final interpretation, authorship, disclosure, and intellectual responsibility remain with the human author. Table 9 indexes the source-free reproducibility package and its integrity checks.

# 17. Conclusion and Future Work

ArchGuard-APR integrates architecture constraints, repository graphs, context retrieval, role-based repair orchestration, patch validation, isolated feedback, and canonical evidence. Phase 13 confirms that these elements can be assembled into auditable research infrastructure, while providing no evidence of repair advantage: both conditions achieved 0/2, all completed outputs were invalid, and two slots attrited before provider use. Future work should prioritize valid-patch production, diagnostic resolution, checkout robustness, and a larger prospectively powered evaluation.

## Data Availability

Canonical source-free data are available under `research/results/phase-13/` and transformed publication data under `paper/data/`. No target source or gold patch is distributed in this package.

## Artifact Availability

The repository provides claim traceability, deterministic build scripts, tables, figures, fingerprints, and verification tests. The target venue remains undecided; the article is not submitted or accepted.

## Author Contributions

The author was solely responsible for conceptualization, methodology, software development, validation, investigation, data curation, formal analysis, visualization, and manuscript preparation.

## Funding

This research received no external funding.

## Conflict of Interest

The author declares no conflict of interest.

# Appendix A. Reproducibility Fingerprints

Final manifest: `8afb87add96e29898ae712f8ef08928b9ba1290a0aabc3ce34bac82adb4b86c2`. Import: `80eac046fc614cad1beb7de127ba69aa907b5c588d5409c30e92e5b685c0dc28`. Population: `421f3129ef6bee5883c41d05f087d6faff5b2a3de3b2ea543b4d27ade1488886`. Protocol: `eae1920ce71c025d2d94096798daf8b14df098249743631efe9a0aa7a541fbe8`. Provider configuration: `f5dc6fb5b8bf54c7cf1b58b990fa3db44073500df9903e3a7d93cbeb1ad92c2d`.